\documentclass[conference]{IEEEtran}

\usepackage{cite}
\usepackage{amsmath}
\usepackage{graphicx}
\usepackage{booktabs}
\usepackage{adjustbox}
\usepackage{algorithm}
\usepackage{algpseudocode}
\usepackage{hyperref}
\hypersetup{
    colorlinks=true,
    linkcolor=blue,
    filecolor=magenta,      
    urlcolor=cyan,
    pdftitle={Spectral Mapping of Singing Voices: U-Net-Assisted Vocal Segmentation},
    pdfpagemode=FullScreen,
    }
\usepackage[section]{placeins}

\begin{document}

\title{Spectral Mapping of Singing Voices: U-Net-Assisted Vocal Segmentation}

\author{
    \IEEEauthorblockN{Adam Sorrenti}
    Toronto Metropolitan University\\
    adam.sorrenti@torontomu.ca
}

\maketitle

\begin{abstract}
Separating vocal elements from musical tracks is a longstanding challenge in audio signal processing. This study tackles the distinct separation of vocal components from musical spectrograms. We employ the Short Time Fourier Transform (STFT) to extract audio waves into detailed frequency-time spectrograms, utilizing the benchmark MUSDB18 dataset for music separation. Subsequently, we implement a UNet neural network to segment the spectrogram image, aiming to delineate and extract singing voice components accurately. We achieved noteworthy results in audio source separation using of our U-Net-based models. The combination of frequency-axis normalization with Min/Max scaling and the Mean Absolute Error (MAE) loss function achieved the highest Source-to-Distortion Ratio (SDR) of 7.1 dB, indicating a high level of accuracy in preserving the quality of the original signal during separation. This setup also recorded impressive Source-to-Interference Ratio (SIR) and Source-to-Artifact Ratio (SAR) scores of 25.2 dB and 7.2 dB, respectively. These values significantly outperformed other configurations, particularly those using Quantile-based normalization or a Mean Squared Error (MSE) loss function. Our source code, model weights, and demo material can be found at the project's GitHub repository: \url{https://github.com/mbrotos/SoundSeg}.
\end{abstract}

\IEEEpeerreviewmaketitle

\section{Introduction}

The field of audio source separation has long been a subject of interest in audio signal processing, driven by its applications in music production, speech recognition, and other audio-related domains. Separating vocal elements from musical tracks is a particularly challenging task within this context, as it involves isolating human singing voices from complex audio mixtures. In this preliminary report and literature review, we examine the topic of spectral mapping of singing voices, specifically focusing on a U-Net \cite{ronneberger2015unet} segmentation architecture. U-Net, initially developed for medical imaging, applies to various areas of signal processing beyond the visual domain. This deep learning architecture has found practical utility in diverse fields, including pictures, medical imaging, and now, audio source separation.

Traditional methods for audio source separation, such as high/low-pass filtering or the use of Deep Neural Networks (DNNs) combined with Principal Component Analysis (PCA) for reducing the dimensionality of spectrogram data \cite{nugraha2016multichannel}, often struggle to effectively isolate singing voices from music. These techniques typically rely on basic spectral manipulations, which can limit their accuracy. In contrast, our study adopts a more sophisticated approach using a U-Net model. We employ the Short Time Fourier Transform (STFT) to transform audio waveforms into detailed frequency-time spectrograms. This approach provides a more comprehensive representation of audio content, aiming to accurately identify and extract the singing voice components from complex audio mixes.

Our investigation employs the well-established MUSDB18 dataset \cite{musdb18}, a benchmark resource for music source separation, to evaluate the performance of our STFT-based U-Net approach. To assess the quality of vocal extraction, we utilize sound source separation metrics such as Signal-to-Distortion Ratios (SDR), Source to Interference Ratios (SIR), and Sources to Artifacts Ratios (SAR). 

Beyond presenting a model for STFT-based source separation performance, this paper explores the robustness of our model across various loss functions. The alignment of data representation and perceptually informed loss functions is an active area of research. The implications of various loss functions and the observed discrepancies between human ratings and standard metrics like SDR is well studied \cite{guso2022loss}. We will discuss empirical differences of different loss functions on audio source separation performance.

The need for convergence on normalization methods is evident in multi-source universal sound separation by integrating semantic embeddings from a pre-trained sound classification network for improved separation, demonstrated by enhanced SDR scores \cite{tzinis2020improving}. We critically examine the impact of Maximum/Minimum and Quantile-based normalization techniques across both time and frequency axes, considering their influence on source separation performance. Additionally, TasNet model \cite{luo2018tasnet} offers an intriguing perspective, despite focusing on time-domain separation, on use of an SDR loss function in an encoder-decoder framework. Our study assesses the suitability of different loss functions within a spectrogram-based framework, contributing to the ongoing discourse in the field of audio source separation.

\section{Problem Statement}
Our goal is to address the challenge of accurately separating singing voices from complex audio mixtures. Thus, we aim to do an initial analysis with a particular focus on examining normalization methods and loss functions using an STFT-based U-Net approach. To achieve this, we will seek answers to the following two \textbf{research questions}~(RQs). 

\textbf{RQ1}. How does the normalization method impact audio source separation?

\textbf{RQ2}. Which loss function is better suited to the task of audio source separation?

The paper makes two major \textbf{contributions}. \textit{First}, we detail the data processing pipeline of STFT-based audio source separation system from waveform to trainable dataset. The data processing pipeline will notably include an initial study into the various normalization methods and will reveal insight into each of their efficacy. Furthermore, we will detail and compare normalization on the time and frequency axis independently. \textit{Second}, we identify two of the most commonly used loss functions for function approximation tasks and apply them in our audio source separation model. We then empirically observe the performance of each normalization and optimization approach across various metrics.

\section{System Model}
Since our STFT-based U-Net approach in audio source separation involves a series of signal processing and deep learning components designed to extract singing voices from mixed audio recordings we first introduce the Short Time Fourier Transform (STFT) at the core of our data pipeline. The transformation will provide the model a time-frequency representation of the input audio signal. The STFT of a time-domain signal \(x(t)\) is defined as:

\begin{equation}
X(\tau, \omega) = \int_{-\infty}^{\infty} x(t) \omega(t - \tau) e^{-i \omega t} \, dt
\end{equation}

Where:
\begin{itemize}
    \item \( X(\tau, \omega) \): Represents phase and magnitude over time and frequency of the signal after transformation.
    \item \( \tau \): This is the time index.
    \item \( \omega(\tau) \): A window function shaping the portion of the signal we analyze. Often a Hann or Gaussian window.
\end{itemize}

The STFT in practice is done in discrete time with quantized variables using the fast Fourier transform. Using the magnitude information we compute the spectrogram representation. This spectrogram is then fed into a U-Net neural network architecture for vocal segmentation.

\begin{figure}[htbp]
\centering
\includegraphics[width=\linewidth]{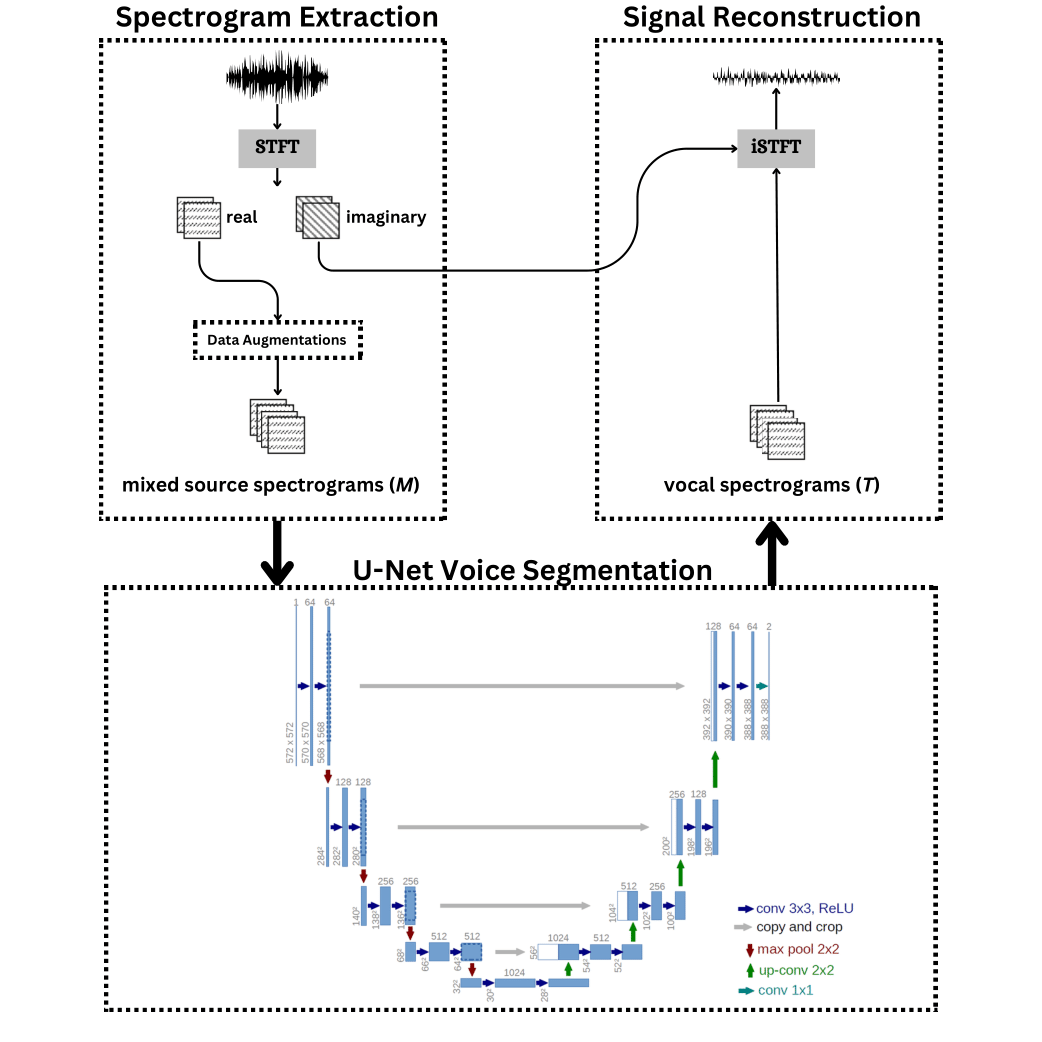}
\caption{A system model depicting the process of complex-valued spectrogram estimation followed by U-Net \cite{ronneberger2015unet} segmentation and subsequent signal reconstruction through iSTFT.}
\label{fig:sysmodel}
\end{figure}

The U-Net architecture consists of two main parts: an encoder and a decoder. The encoder captures high-level features from the input spectrogram, while the decoder generates a segmented spectrogram that contains the singing voice components. The data pipeline and U-Net operations is represent in Figure~\ref{fig:sysmodel}.

The U-Net learns through regression to approximate the singing voice components of the input mixture. A mask can also be generated by taking the ratio between the voice and mixture spectrogram components. The mask can be inverted and used to isolate the accompaniments of instrumental components of the input mixture. Finally, the resulting spectrogram isolate can be converted back to waveform using the inverse Short Time Fourier Transform (iSTFT) and the original complex-valued phase information.

\subsection{Data preprocessing}

To prepare the spectral data for training, it was reshaped into uniform training examples. Initially, the waveform data from the MUSDB dataset was downsampled from the original 44.1 kHz to 11 kHz. This preprocessing step was performed using the audio and music processing toolkit Librosa \cite{mcfee_2023_8252662}. The downsampled waveform data was then transformed using the Short Time Fourier Transform (STFT) with a frame size of 1024---a power of two to optimize the Fast Fourier Transform (FFT) algorithm---a hop length of 256, and a Hann window function. The resulting complex-valued spectrogram \( \mathbf{D} \) was decomposed into its magnitude (\( \mathbf{S} \)) and phase (\( \mathbf{P} \)) components. These components were subsequently reshaped into dimensions of \( (n, 512, 128) \), where \( n \) represents the number of spectral samples, with each sample having 512 frequencies over 128 time steps. The values within the spectrogram samples indicate the intensity of a given frequency at each time step.\footnote{The STFT with a frame size of 1024 produces 513 frequency representations. The highest frequency component is typically discarded for convenience, resulting in 512 frequency components.}

\subsection{Data Augmentations}

\begin{figure}[!htb]
\centering
\includegraphics[width=\linewidth]{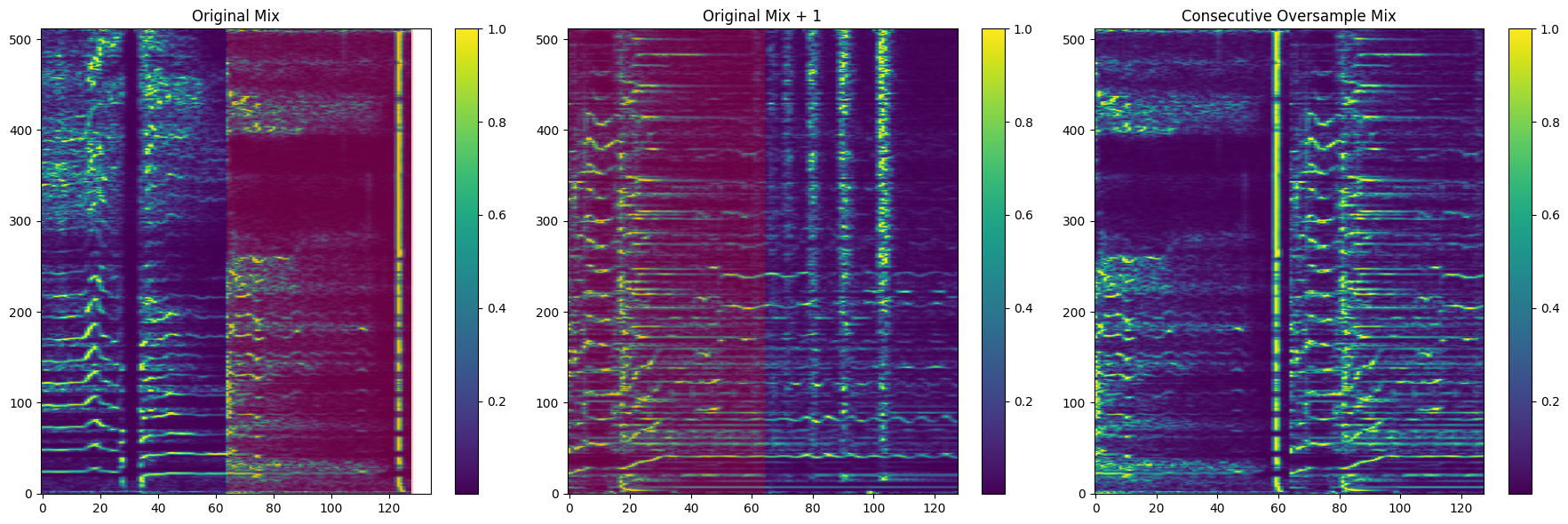}
\caption{Spectrogram examples illustrating the effect of splicing. The leftmost image shows the original mix, the middle image shows the original mix shifted by one time step, and the rightmost image presents the result of splicing the mix spectrograms. The red highlights show the component halves.}
\label{fig:splicing}
\end{figure}

\begin{figure}[!htb]
\centering
\includegraphics[width=0.9\linewidth]{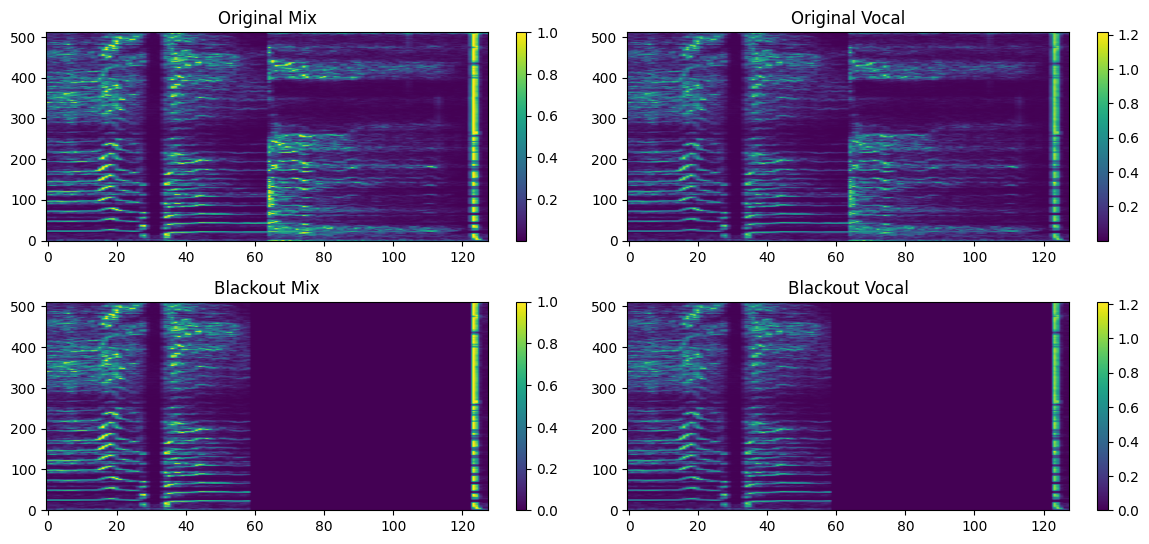}
\caption{Spectrogram examples showing the blackout data augmentation. The top row displays the original mix and vocal spectrograms, while the bottom row demonstrates the effect of the blackout augmentation on both.}
\label{fig:blackout}
\end{figure}

In our study, we employed two primary data augmentation techniques for audio spectrogram processing. The first technique, described in Algorithm~\ref{alg:splicing} and visually represented in Figure~\ref{fig:splicing}, involves consecutive oversampling of spectrograms to generate more training data. This method concatenates halves of adjacent spectrograms along the time axis. The second technique, as outlined in Algorithm~\ref{alg:blackout} and illustrated in Figure~\ref{fig:blackout}, applies a 'blackout' operation on both mix and vocal spectrograms. This process randomly zeroes out segments along the time axis for both types of spectrograms, enhancing the robustness of our model, as detailed in the algorithm and visually exemplified in the corresponding figure.

\begin{algorithm}[!htb]
\caption{Splicing Augmentation}
\label{alg:splicing}
\begin{algorithmic}[1] 
\Require $mixes$:  Mixture spectrograms (None, 512, 128, 1)
\Require $vocals$: Vocal spectrograms (None, 512, 128, 1)
\Ensure Concatenated spectrograms of mixes and vocals
\newline
\Function{ConcatenateHalves}{$data$}
    \State $half\_size \gets$ Third dimension of $data$ divided by 2
    \For{$i = 0$ \textbf{to} $len(data) - 2$}
        \State $first \gets$ Last $half\_size$ columns of $data[i]$
        \State $second \gets$ First $half\_size$ columns of $data[i+1]$
    \EndFor \Return \newline Concatenate $first\_half$ and $second\_half$ along axis 2
\EndFunction
\newline
\end{algorithmic}
\end{algorithm}

\begin{algorithm}[!htb]
\caption{Blackout Augmentation}
\label{alg:blackout}
\begin{algorithmic}[1] 
\Require $mixes$:  Mixture spectrograms (None, 512, 128, 1)
\Require $vocals$: Vocal spectrograms (None, 512, 128, 1)
\Ensure Blackout applied spectrograms of mixes and vocals
\newline
\State $blackout\_size \gets$ 64
\For{$i = 0$ \textbf{to} $len(mixes) - 1$}
    \State $start \gets$ Random integer from 0 to 64
    \State Set frequencies to zero in \newline $mix\_blackout[i, :, start:start + blackout\_size, :]$
    \State Set frequencies to zero in \newline $vocal\_blackout[i, :, start:start + blackout\_size, :]$
\EndFor
\Return $mix\_blackout$, $vocal\_blackout$
\end{algorithmic}
\end{algorithm}

\subsection{Normalization}

\begin{figure}[!htb]
\centering
\includegraphics[width=0.6\linewidth]{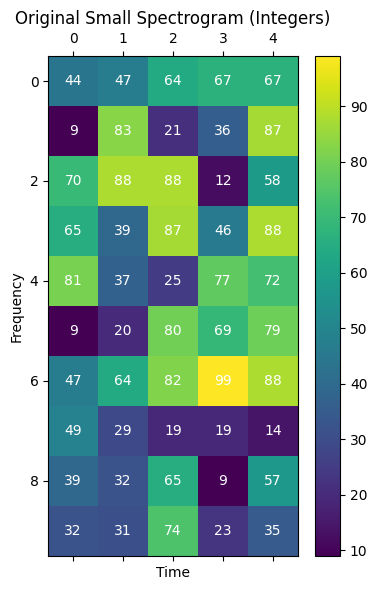}
\caption{The original example integer spectrogram before normalization with Time on the x-axis and Frequency on the y-axis. The intensity values were randomly assigned.}
\label{fig:original}
\end{figure}

\begin{figure}[!htb]
\centering
\includegraphics[width=\linewidth]{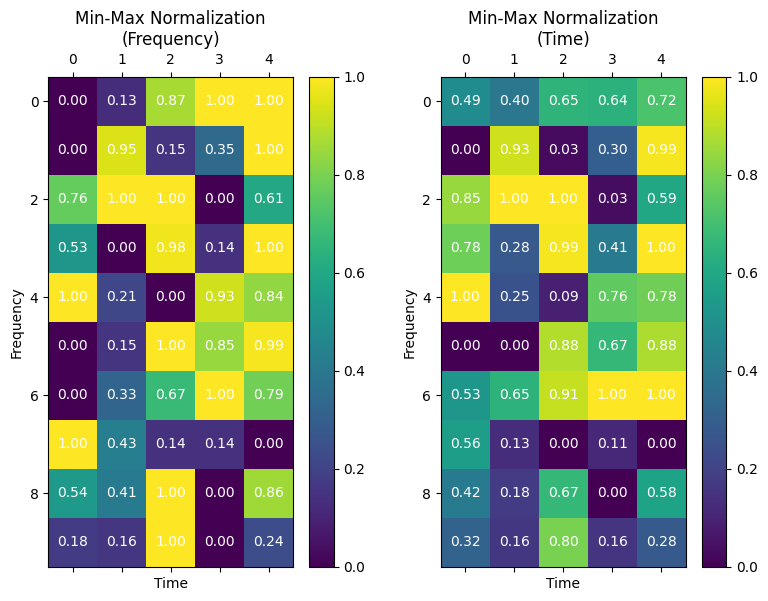}
\caption{Min/max normalization applied to the spectrogram. The left plot shows min/max normalization across the frequency axis, while the right plot applies min/max normalization across the time axis.}
\label{fig:min_max_normalization}
\end{figure}

In our data prepossessing steps for audio source separation, we employed two distinct normalization techniques on spectrogram data, each applied separately to the frequency and time axes. This approach allows us to analyze how different normalization methods impact the performance of audio source separation models. Given an input spectrogram, as show in Figure~\ref{fig:original} by a small integer-based example, we applied various normalization methods. Depicted in Figure~\ref{fig:min_max_normalization}, min/max normalization linearly scales the data between the minimum and maximum values for each feature. For a given feature \( x \), min/max normalization is computed as:

\[ x_{\text{norm}} = \frac{x - x_{\text{min}}}{x_{\text{max}} - x_{\text{min}}} \]

\begin{figure}[!htb]
\centering
\includegraphics[width=\linewidth]{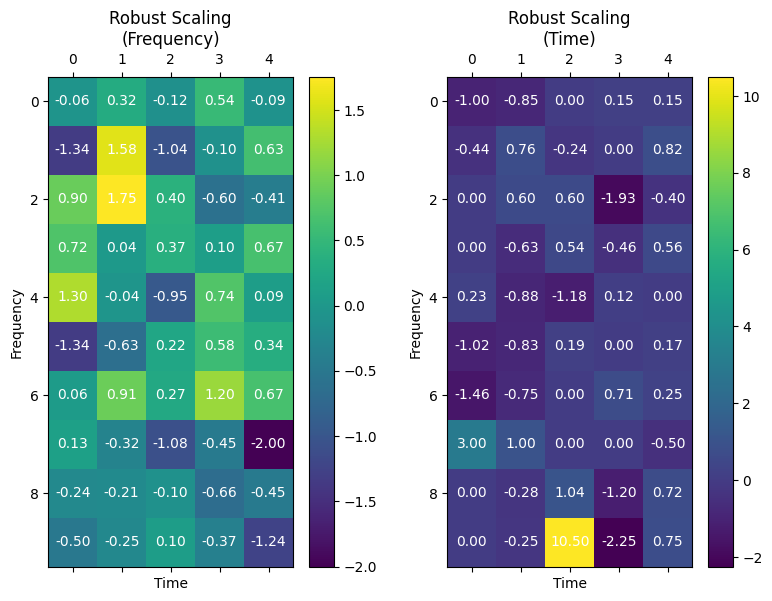}
\caption{Robust scaling applied to the spectrogram. The left plot shows robust scaling across the frequency axis, while the right plot applies robust scaling across the time axis.}
\label{fig:robust_scaling}
\end{figure}

This technique was applied independently to each frequency bin and each time step to observe its effects on different dimensions of the spectrogram data. Similarly, robust scaling, illustrated in Figure~\ref{fig:robust_scaling}, employs the median and the interquartile range (IQR) to scale features, offering enhanced resistance to outliers. Robust scaling for a feature \( x \) is defined as:

\[ x_{\text{robust}} = \frac{x - \text{median}(x)}{\text{IQR}(x)} \]

where \( \text{IQR}(x) \) is the range between the 25th and 75th percentiles of \( x \). This method was also applied separately along the frequency and time axes using a popular implementation in the Scikit-learn package \cite{scikit-learn}.

\subsection{Evaluation Metrics}

Evaluating source separation methods is complex and typically involves two types of assessments: objective and subjective. Objective evaluations use calculations to compare the output of the separation system against known isolated sources, focusing on measurable qualities. In contrast, subjective evaluations rely on human listeners who rate the system's output based on their perception. While objective methods are quicker and more cost-effective, they struggle to fully capture the nuances of human auditory perception. Subjective methods, though more time-consuming and variable due to human involvement, can offer more reliable insights into the actual listening experience. Our study will focus on the main three objective measures of source separation performance as implemented in \textit{mir\_eval}, a popular Music Information Retrieval toolkit \cite{mir_eval}.

Currently, the most prevalent methods for assessing the performance of a source separation system are the Source-to-Distortion Ratio (SDR), Source-to-Interference Ratio (SIR), and Source-to-Artifact Ratio (SAR).

An estimated source is represented as $\hat{s}_i$, this estimate is assumed to consist of four distinct components:

$$
\hat{s}_i = s_{\text{target}} + e_{\text{interf}} + e_{\text{noise}} + e_{\text{artif}},
$$

Here, $s_{\text{target}}$ denotes the authentic source, while $e_{\text{interf}}$, $e_{\text{noise}}$, and $e_{\text{artif}}$ represent the errors due to interference, noise, and introduced artifacts, respectively \cite{vincent2006performance}.

These four components are utilized to define our metrics. All these metrics are expressed in decibels (dB), where higher values indicate superior performance. Their calculation requires the original isolated sources and is typically performed on signals segmented into brief time frames, usually a few seconds each.

\subsubsection{Source-to-Distortion Ratio (SDR)}
$$
\text{SDR} := 10 \log_{10} \left( \frac{\| s_{\text{target}} \|^2}{ \| e_{\text{interf}} + e_{\text{noise}} + e_{\text{artif}} \|^2} \right)
$$

SDR is a measure of the amount of distortion introduced to a signal compared to the original source signal. A higher SDR value indicates that the separated signal is closer to the original source signal, meaning less distortion. In other words, SDR assesses how well the source separation process has preserved the quality of the original signal while isolating it from other sources.

\subsubsection{Source-to-Interference Ratio (SIR)}
$$
\text{SIR} := 10 \log_{10} \left( \frac{\| s_{\text{target}} \|^2}{ \| e_{\text{interf}} \|^2} \right)
$$

SIR measures the level of interference from other sources in the separated signal. A higher SIR value indicates that the separated signal contains less interference from other sources. SIR is particularly important in scenarios where multiple sources are present, such as in music with several instruments or in environments with multiple speakers. 

\subsubsection{Source-to-Artifact Ratio (SAR)}
$$
\text{SAR} := 10 \log_{10} \left( \frac{\| s_{\text{target}} + e_{\text{interf}} + e_{\text{noise}} \|^2}{ \| e_{\text{artif}} \|^2} \right)
$$

SAR quantifies the amount of artifacts, or unwanted alterations, introduced to the signal by the source separation process. Artifacts can include noises, echoes, or other distortions that were not present in the original signal. A higher SAR value suggests fewer artifacts, indicating a cleaner separation process.

\section{Experimental Setup}

Our experimental setup focuses on evaluating the impact of different normalization methods and loss functions on the performance of a U-Net-based audio source separation model. The experiments are structured to systematically assess these factors across various combinations. The key components of our experimental setup include the dataset, model architecture, training parameters, and the specific combinations of hyperparameters we tested.

\subsection{Dataset}

The dataset used in this study is the MUSDB18 dataset \cite{musdb18}, a widely recognized benchmark in the field of music source separation. It consists of a collection of music tracks with separate stems for vocals, drums, bass, and other instruments. For our experiments, we focus on the task of separating singing voices from the mixed tracks. The dataset includes 75 tracks for training and 75 tracks for testing, with a variety of genres represented.

\subsection{Model Architecture}

Our model is based on the U-Net architecture \cite{ronneberger2015unet}, modified for audio source separation tasks. The U-Net model consists of an encoder-decoder structure with skip connections, allowing for the capture of both low-level and high-level features of the audio spectrograms.

\begin{itemize}
    \item \textbf{Encoder:} The encoder comprises a series of Convolutional and MaxPooling layers, which reduce the spatial dimensions of the input while increasing the feature dimensions.
    \item \textbf{Decoder:} The decoder uses Convolutional Transpose layers to upsample the feature maps, concatenated with corresponding feature maps from the encoder via skip connections.
    \item \textbf{Output:} The final layer of the model is a Conv1D layer with a linear activation function, producing the estimated vocal magnitude spectrogram.
\end{itemize}

\subsection{Training Parameters}

The model is trained using the following parameters:

\begin{itemize}
    \item \textbf{Batch Size:} 64
    \item \textbf{Epochs:} 20
    \item \textbf{Optimizer:} Adam
    \item \textbf{Learning Rate:} 0.001
    \item \textbf{Loss Functions:} Mean Squared Error (MSE) and Mean Absolute Error (MAE)
\end{itemize}

\subsection{Hyperparameter Combinations}

We trained models for the 2x2x2 combinations of the following hyperparameters:

\begin{itemize}
    \item \textbf{Normalization Axis:} Time, Frequency
    \item \textbf{Loss Function:} Mean Absolute Error (MAE), Mean Squared Error (MSE)
    \item \textbf{Scaler:} Min/Max, Quantile/Robust
\end{itemize}

These combinations allow us to analyze the impact of each factor on the model's performance in isolating singing voices from music tracks.

\subsection{Model Selection and Validation}

For each combination of hyperparameters, the best-performing model was selected based on the minimum validation loss observed at the end of the training epochs. The validation dataset constituted 10\% of the total training dataset, providing a separate and unbiased evaluation of the model performance. This approach ensures that the chosen model has not only learned to generalize well but also avoided overfitting to the training data.

\subsection{Training Dataset Composition}

The training dataset comprised 9141 pairs of (512, 128, 1) spectrogram mixture-vocal examples. These pairs were derived from the MUSDB18 dataset and processed through our STFT-based pipeline to generate the appropriate spectrogram representations. The composition of the training dataset was as follows:

\begin{itemize}
    \item \textbf{Original Examples:} 40\% of the training data consisted of unmodified spectrogram mixture-vocal pairs directly obtained from the MUSDB18 dataset.
    \item \textbf{Splicing Augmentation:} Another 40\% of the training data included examples generated through the splicing augmentation technique. This method involved concatenating halves of adjacent spectrograms along the time axis, as described earlier, to increase data diversity and robustness.
    \item \textbf{Blackout Augmentation:} The remaining 20\% of the training data were created using the blackout augmentation technique. This process involved randomly zeroing out segments of both mix and vocal spectrograms along the time axis, further enhancing the model's ability to handle a variety of audio scenarios.
\end{itemize}

These augmentations were crucial in expanding the diversity of the training data, thereby enabling the model to learn more generalized features and perform more robustly on unseen data.

\section{Results and Discussion}

In our evaluation of the U-Net-based models for audio source separation, the experimental results detailed in Table~\ref{table:results} provide a comprehensive understanding of the effects of different normalization methods and loss functions on the separation quality.

\begin{table}[htbp]
\centering
\caption{Results of the U-Net-based models for audio source separation.}
\label{table:results}
\begin{tabular}{rrrlrl}
\toprule
SDR & SIR & SAR & Normalization & Scaler & Loss \\
\midrule
\textbf{7.1} & \textbf{25.2} & \textbf{7.2} & frequency & Min/Max & MAE \\
\textbf{7.1} & 25.1 & \textbf{7.2} & time & Min/Max & MAE \\
6.7 & 24.8 & 6.8 & frequency & Min/Max & MSE \\
5.7 & 23.9 & 5.8 & time & Quantile & MAE \\
5.6 & 23.3 & 5.7 & time & Min/Max & MSE \\
4.8 & 22.6 & 4.9 & time & Quantile & MSE \\
-0.9 & 16.6 & -0.6 & frequency & Quantile & MSE \\
-2.1 & 15.8 & -1.8 & frequency & Quantile & MAE \\
\bottomrule
\end{tabular}
\end{table}

\subsection{Normalization and Scaling Function}

Regarding RQ1, the axis of normalization and the type of scaler used were observed to have a significant influence on the model's performance. As shown in Table~\ref{table:results}, the models with Min/Max scaling, regardless of the normalization axis, achieved the highest scores in all three metrics (SDR, SIR, and SAR), with the frequency normalization paired with MAE loss function, achieving the top score of \textbf{7.1 dB} in SDR. This suggests that Min/Max scaling is better suited to this task, possibly due to its ability to preserve the original distribution of the frequency intensities.

\subsection{Loss Function Analysis}

In addressing RQ2, the Mean Absolute Error (MAE) loss function emerged superior to Mean Squared Error (MSE), as evidenced by its consistent presence in the top-performing models. Models employing MAE achieved the highest SDR and SAR scores, which could be attributed to MAE's characteristic of not disproportionately penalizing larger errors, potentially beneficial in handling the variability inherent in audio signals.

\subsection{Comprehensive Evaluation and Implications}

When considering the overall performance, it is evident from Table~\ref{table:results} that the combination of Min/Max scaling with MAE loss provides a robust approach for audio source separation. This is particularly important when considering the separation of a complex audio signal where most frequencies are irrelevant to the task at hand, and large errors in estimating the intensity of frequencies can occur. The MAE loss function appears to manage these errors without significantly compromising the quality of the audio separation.

\subsection{Conclusion and Future Work}

The results conclusively demonstrate the effectiveness of Min/Max scaling in conjunction with the MAE loss function for the task of audio source separation. In future work, it would be worthwhile to explore other loss functions, such as an SDR-based loss, which may offer a more direct optimization path for the quality metrics used in this field. Additionally, the augmentation strategies can be enhanced by introducing static tones at different frequencies or adding varying levels of noise, which could further improve the robustness and generalizability of the separation model.

In conclusion, our experiments indicate a clear path forward for improving audio source separation models. By focusing on the scaling method and loss function, we can significantly enhance the performance of these models, leading to better quality separations that could benefit a wide range of applications in the audio industry.

\bibliographystyle{IEEEtran}
\bibliography{bib}

\begin{thebibliography}{10}
\providecommand{\url}[1]{#1}
\csname url@samestyle\endcsname
\providecommand{\newblock}{\relax}
\providecommand{\bibinfo}[2]{#2}
\providecommand{\BIBentrySTDinterwordspacing}{\spaceskip=0pt\relax}
\providecommand{\BIBentryALTinterwordstretchfactor}{4}
\providecommand{\BIBentryALTinterwordspacing}{\spaceskip=\fontdimen2\font plus
\BIBentryALTinterwordstretchfactor\fontdimen3\font minus \fontdimen4\font\relax}
\providecommand{\BIBforeignlanguage}[2]{{%
\expandafter\ifx\csname l@#1\endcsname\relax
\typeout{** WARNING: IEEEtran.bst: No hyphenation pattern has been}%
\typeout{** loaded for the language `#1'. Using the pattern for}%
\typeout{** the default language instead.}%
\else
\language=\csname l@#1\endcsname
\fi
#2}}
\providecommand{\BIBdecl}{\relax}
\BIBdecl

\bibitem{ronneberger2015unet}
O.~Ronneberger, P.~Fischer, and T.~Brox, ``U-net: Convolutional networks for biomedical image segmentation,'' 2015.

\bibitem{nugraha2016multichannel}
A.~A. Nugraha, A.~Liutkus, and E.~Vincent, ``Multichannel audio source separation with deep neural networks,'' \emph{IEEE/ACM Transactions on Audio, Speech, and Language Processing}, vol.~24, no.~9, pp. 1652--1664, 2016.

\bibitem{musdb18}
\BIBentryALTinterwordspacing
Z.~Rafii, A.~Liutkus, F.-R. St{\"o}ter, S.~I. Mimilakis, and R.~Bittner, ``The {MUSDB18} corpus for music separation,'' Dec. 2017. [Online]. Available: \url{https://doi.org/10.5281/zenodo.1117372}
\BIBentrySTDinterwordspacing

\bibitem{guso2022loss}
E.~Gus{\'o}, J.~Pons, S.~Pascual, and J.~Serr{\`a}, ``On loss functions and evaluation metrics for music source separation,'' in \emph{ICASSP 2022-2022 IEEE International Conference on Acoustics, Speech and Signal Processing (ICASSP)}.\hskip 1em plus 0.5em minus 0.4em\relax IEEE, 2022, pp. 306--310.

\bibitem{tzinis2020improving}
E.~Tzinis, S.~Wisdom, J.~R. Hershey, A.~Jansen, and D.~P. Ellis, ``Improving universal sound separation using sound classification,'' in \emph{ICASSP 2020-2020 IEEE International Conference on Acoustics, Speech and Signal Processing (ICASSP)}.\hskip 1em plus 0.5em minus 0.4em\relax IEEE, 2020, pp. 96--100.

\bibitem{luo2018tasnet}
Y.~Luo and N.~Mesgarani, ``Tasnet: time-domain audio separation network for real-time, single-channel speech separation,'' in \emph{2018 IEEE International Conference on Acoustics, Speech and Signal Processing (ICASSP)}.\hskip 1em plus 0.5em minus 0.4em\relax IEEE, 2018, pp. 696--700.

\bibitem{mcfee_2023_8252662}
\BIBentryALTinterwordspacing
M.~et~al, ``librosa/librosa: 0.10.1,'' Aug. 2023. [Online]. Available: \url{https://doi.org/10.5281/zenodo.8252662}
\BIBentrySTDinterwordspacing

\bibitem{scikit-learn}
\BIBentryALTinterwordspacing
F.~Pedregosa, G.~Varoquaux, A.~Gramfort, V.~Michel, B.~Thirion, O.~Grisel, M.~Blondel, P.~Prettenhofer, R.~Weiss, V.~Dubourg, J.~Vanderplas, A.~Passos, D.~Cournapeau, M.~Brucher, M.~Perrot, and E.~Duchesnay, ``Scikit-learn: Machine learning in {P}ython,'' \emph{Journal of Machine Learning Research}, vol.~12, pp. 2825--2830, 2011. [Online]. Available: \url{https://scikit-learn.org/stable/modules/generated/sklearn.preprocessing.RobustScaler.html}
\BIBentrySTDinterwordspacing

\bibitem{mir_eval}
\BIBentryALTinterwordspacing
C.~Raffel, B.~Mcfee, E.~Humphrey, J.~Salamon, O.~Nieto, D.~Liang, and D.~Ellis, ``mir\_eval: A transparent implementation of common mir metrics,'' in \emph{Proceedings - 15th International Society for Music Information Retrieval Conference (ISMIR 2014)}, 10 2014. [Online]. Available: \url{https://github.com/craffel/mir_eval}
\BIBentrySTDinterwordspacing

\bibitem{vincent2006performance}
E.~Vincent, R.~Gribonval, and C.~Fevotte, ``Performance measurement in blind audio source separation,'' \emph{IEEE Transactions on Audio, Speech, and Language Processing}, vol.~14, no.~4, pp. 1462--1469, 2006.

\end{thebibliography}

\end{document}